\documentstyle[12pt]{article}
\include{epsf}
\begin{document}

\thispagestyle{empty}

\begin{flushright}
UNIGRAZ-UTP-14-07-97 \\
hep-lat/9707011
\end{flushright}
\begin{center}
\vspace*{5mm}
{\Large Topological Charge and the Spectrum of \vskip2mm
the Fermion Matrix in Lattice-QED$_2{\; }^*$}
\vskip9mm
\centerline{ {\bf
C.R. Gattringer}}
\vskip 2mm
\centerline{Department of Physics and Astronomy,}
\centerline{University of British Columbia, Vancouver B.C., Canada}
\vskip 5mm
\centerline{ {\bf
I. Hip and C.B. Lang }}
\vskip2mm
\centerline{Institut f\"{u}r Theoretische Physik} 
\centerline{Universit\"at Graz, A-8010 Graz, Austria}
\end{center}
\vskip5mm
\begin{abstract}
We investigate the interplay between topological charge and the spectrum of 
the fermion matrix in lattice-QED$_2$ using analytic methods and 
Monte Carlo simulations with dynamical fermions. A new theorem 
on the spectral decomposition
of the fermion matrix establishes that its real eigenvalues 
(and corresponding eigenvectors) play a
role similar to the zero eigenvalues (zero modes) of the Dirac operator
in continuous background fields. Using numerical techniques we 
concentrate on studying the real part of the spectrum. These results 
provide new insights into the behaviour of physical quantities as a
function of the topological charge. In particular we discuss 
fermion determinant, effective action and pseudoscalar densities.
\end{abstract}
\vskip3mm \nopagebreak \begin{flushleft} \rule{2 in}{0.03cm}
\\ {\footnotesize \ 
${}^*$ Supported by Fonds zur F\"orderung der Wissenschaftlichen 
Forschung in \"Osterreich, Projects P11502-PHY and J01185-PHY.}
\end{flushleft}
\newpage
\setcounter{page}{1}
\section{Introduction}

Topological concepts play a prominent role in our understanding of 
continuum gauge theories. An example of particular importance
is the idea that for the true vacuum functional of QCD 
($\theta$-vacuum) gauge field configurations of all topological charges 
have to be taken into account 
in the path integral \cite{thetavac}. Such statements however 
are difficult to 
formulate in a mathematically precise way. The topological charge is a 
concept for classical, i.e.~differentiable fields, since only those fields 
can be uniquely
classified with respect to an integer topological charge. On the other 
hand differentiable configurations are of measure zero in the path integral 
for the gauge fields (see e.g.~\cite{CoLa73}). 
 
When formulating gauge theories on the lattice the notion of a 
classical gauge field configuration gets lost. However, various
definitions of a topological charge for lattice gauge fields have 
been proposed (see e.g.~the review \cite{Kr88} and also
\cite{Gi96} for an overview on more recent developments). 
Using such a prescription one can in principle assign a 
topological charge to 
{\sl every} lattice configuration (up to so-called
exceptional configurations which in the continuum limit do not contribute 
to the path integral). In a certain sense this makes the 
lattice approach very powerful for pursuing topological concepts since
the path integral can indeed be decomposed into topological sectors
whereas in the continuum formulation topological arguments are restricted 
to (non-contributing) classical configurations. 

The working hypothesis, that every gauge field configuration contributing
to the lattice path integral in the continuum limit can be assigned a topological charge, allows to decompose the lattice path integral
into topological sectors. As discussed above, in the continuum 
this is only possible in a formal way. The topological decomposition
of the {\sl lattice} path integral is a conceptionally sound procedure
and allows to address many interesting physical questions in a 
mathematically meaningful setting.
\\

For classical gauge field configurations the Atiyah Singer Index Theorem 
\cite{AtSi71} provides
a powerful tool connecting the topological charge with the zero 
modes of the Dirac operator. However in the continuum path integral 
its value is reduced to formal or semi-classical arguments, since the
classical configurations are of measure zero as remarked above.
 
On the lattice the situation is different. On one hand 
when approaching the continuum limit, every gauge field 
configuration contributing in the continuum limit can be assigned a 
topological charge. On the other hand we have no such thing as the
Atiyah Singer Index Theorem on the lattice. However, in the literature 
several investigations \cite{KaSeSt86} - \cite{BaDuEiTh97} attempting to 
establish the Index Theorem on the lattice in a probabilistic sense 
can be found. If such a goal is achieved, together with the fact that
the lattice path integral can be decomposed into topological sectors, 
this would allow to gain better insight into the behaviour of fermionic
observables using the connection between topology and the spectrum of
the fermion matrix.
\\

In this article we study QED$_2$ on the lattice to demonstrate that
a tight connection between topology and spectrum of the fermion matrix 
can be found and provides a powerful mathematical tool. 
Establishing this connection 
is a two step procedure. In a first step we use analytic arguments 
to understand general features of the spectrum of the fermion matrix. 
In particular we prove a new theorem on the chiral properties
of the eigenvectors of the fermion matrix, which makes clear 
that only the real eigenvalues play a special role similar to 
zero eigenvalues (and zero modes) of the continuum Dirac operator. 
These analytic results can be generalized to the 4-dimensional 
Wilson-Dirac operator \cite{Ga97}.

Once the real eigenvalues are identified as being the trace of 
the topological charge in the spectrum, in a second step we 
concentrate on analyzing their properties using numerical 
methods. Tracing only the real eigenvalues is a much simpler task 
than investigating all eigenvalues of the fermion matrix. 
For sufficiently large $\beta$ 
we establish results on size and distribution of the real eigenvalues
and show that their number is related to the topological charge
computed with the geometric definition \cite{Lu82}.
In a final section we make use of these results and use them to 
discuss the behaviour of the fermion determinant, the effective action
and the pseudoscalar densities. 
\\

The model under consideration (QED$_2$) resembles many features of 
4-dimensional gauge theories such as QCD. In particular U(1) gauge theory 
in 2 dimensions allows for classical topologically nontrivial 
configurations (`vortices') 
which play the role of the instantons of 4D Yang-Mills theory. QED$_2$
has an anomaly which is related to the mass generation of 
the pseudoscalar sing\-let as in QCD. Thus the lattice version of 
this low dimensional model is an interesting candidate when exploring 
the role of topology in lattice models.\- Moreover for the continuum 
model there exists a limit which can be solved analytically. When setting 
the fermion masses to zero, the model reduces to the Schwinger 
model \cite{Sch62} for two species of fermions. This allows to check the 
lattice simulations and to
gain intuition from the analytic result. For the case of massive fermions
semiclassical results \cite{masch} and expansions in the bare mass
are available \cite{mexp}. 
\\

The article is organized as follows: Sec.~2 contains 
a short discussion of the model, the details of the simulation as well as
definitions and theorems for the topological charge in the continuum and on 
the lattice. In Sec.~3 we develop the analytic results for the 
spectral decomposition of the fermion matrix and in Sec.~4 we 
analyze the spectrum using numerical techniques. Finally in Sec.~5 
we discuss the applications outlined above.
The article ends with a discussion (Sec.~6).

\section{Setting}
This section prepares the ground for the program outlined in the 
introduction. We formulate the lattice model, present technical details of 
the simulation and discuss the definitions and results for the 
topological charge in the continuum as well as on the lattice. 
 
\subsection{QED$_2$ on the Lattice}
We work on a two dimensional lattice $\Lambda$ with volume $L^2$. 
Lattice sites are denoted
as $x = (x_1,x_2)$ with $x_i = 1,2,...,L$. The lattice spacing is set
equal to 1. The gauge fields
are group elements $U_\nu(x) \in \mbox{U(1)}$ assigned to the 
links between nearest neighbours $x,x+\hat{\nu}$ and their action is given by
\begin{equation}
S_g \; = \; \beta \sum_{x \in \Lambda} \;
\Big[ \; 1  \; - \; \mbox{Re} \; U_P(x) \; \Big] \; ,
\label{latact}
\end{equation}
where the plaquette element is defined as
\begin{equation}
U_P(x) \; = \; U_1(x) \; U_2(x+\hat{1}) \;
\overline{U_1(x+\hat{2})} \;\overline{U_2(x)} \; .
\label{plaquel}
\end{equation}
The gauge fields obey periodic boundary conditions $U_\nu(L+1,x_2) = 
U_\nu(1,x_2)$, $U_\nu(x_1,L+1) = U_\nu (x_1,1)$.
The fermion action is the bilinear form (in matrix notation)
\[
S_f \; = \; -\overline{\psi} M \psi \; ,
\]
where the (implicit) summation is over lattice points and spinor indices - 
the above mentioned two flavors come about through squaring the fermion
matrix (see below). We write the fermion matrix as
\begin{equation}
M \; = \; 1 - \kappa Q \; ,
\label{mq}
\end{equation}
where the hopping matrix $Q$ is defined as (spinor indices still suppressed) 
\begin{eqnarray}
Q (x,y)  &=&  \sum_{\nu = 1,2} Q_\nu(x,y)\;,\nonumber\\
Q_\nu(x,y)&=&
(1+\gamma_\nu) U_\nu (x - \hat{\nu}) \Delta_{x-\hat{\nu},y} \; + \;
(1 - \gamma_\nu) \overline{U_\nu (x)}\Delta_{x+\hat{\nu},y}\; .
\label{Qferm} 
\end{eqnarray}
$\beta$ is related to the bare coupling constant $e$ via 
$\beta = 1/e^2$ and the hopping parameter $\kappa$ to the bare mass $m$
through $\kappa = (2m + 4)^{-1}$. The matrices $\gamma_\nu$ are chosen as
the Pauli matrices $\gamma_\nu = \sigma_\nu, \; \nu = 1,2,3$. The fermions 
obey mixed boundary conditions, i.e. periodic in $\hat{1}$-direction and 
antiperiodic in $\hat{2}$-direction. 
This is taken into account by the mixed 
periodic Kronecker delta defined as
\begin{eqnarray}
\Delta_{x-\hat{1},y} & \; = \; &
\delta_{x_2,y_2} \Big[ \delta_{x_1,1} \delta_{y_1,L} + 
\delta_{x_1,2} \delta_{y_1,1} + .... \; \; \;  
\delta_{x_1,L} \delta_{y_1,L-1} \Big] \; ,
\nonumber \\
\Delta_{x+\hat{1},y} & \; = \; &
\delta_{x_2,y_2} \Big[ \delta_{x_1,1} \delta_{y_1,2} + .... \; \; \; 
\delta_{x_1,L-1} \delta_{y_1,L} + \delta_{x_1,L} \delta_{y_1,1} \Big] \; ,
\nonumber \\
\Delta_{x-\hat{2},y} & \; = \; &
\delta_{x_1,y_1} \Big[ - \delta_{x_2,1} \delta_{y_2,L} + 
\delta_{x_2,2} \delta_{y_2,1} + .... \; \; \; 
\delta_{x_2,L} \delta_{y_2,L-1} \Big] \; ,
\nonumber \\
\Delta_{x+\hat{2},y} & \; = \; &
\delta_{x_1,y_1} \Big[ \delta_{x_2,1} \delta_{y_2,2} + .... \; \; \; 
\delta_{x_2,L-1} \delta_{y_2,L} - \delta_{x_2,L} \delta_{y_2,1} \Big] \; .
\label{kron}
\end{eqnarray}
Obviously $\Delta_{y-\hat{\nu},x} \; = \; \Delta_{x+\hat{\nu},y}$.
We are rather explicit here, since the analytic results for the spectrum 
of the fermion matrix in a fixed background configuration depend on the
boundary conditions and also change whether $L$ is even or odd. The
latter dependence can be discussed conveniently using 
$\Delta_{x \pm \hat{\nu}, y}$. It has to be remarked, that when 
actually integrating
over all gauge configurations the boundary conditions for the fermions 
are irrelevant
for this model, since gauge group U(1) contains also $-1$.
For a given gauge field configuration there is a partner 
differing just in the boundary links by a factor $-1$. The 
operators resulting from the Grassmann integral over fermion
thus are averaged over these configurations.

\subsection{Technical details of the simulation}

We use lattices of size $L^2$ with $L$ between 4 and 16. In the Monte
Carlo simulation the fermions were considered with the hybrid Monte
Carlo method \cite{DuKePe87}. In order to ensure positivity of the
fermion determinant measure we had to introduce a second species of
fermions. We stress that we squared the fermion determinant only for
generating Monte Carlo configurations, whereas below we analyze the
spectrum and other properties of $M$ as defined in (\ref{mq}) and
(\ref{Qferm}) (not $M^2$).  In the HMC method we used trajectories with
10 steps and a step size adjusted such that the acceptance rate in the 
Monte Carlo step  was 0.8 in the average.

For the inversion of the fermion determinant even-odd preconditioning
was used, together with the BiCG$\gamma_5$ (actually BiCG$\sigma_3$)
algorithm \cite{Fo95}.  This proved to be efficient for most
configurations.  Whenever we found slow convergence (so-called
``unstable configurations'', also often called exceptional, in
contradistinction to the notation for configurations without definite
topological charge to be discussed later) we switched the inverter and
used BiCGStab (biconjugate gradient stabilized, cf. e.g. \cite{FrHa94})
without preconditioning, which seemed to be the most reliable
(although not most efficient) method.

In general we do not observe any abundance of unstable configurations
close to $\kappa_{crit}$ but actually only very few (${\cal O}(0.1\%)$),
which then were successfully inverted by BiCGStab.  
In fact, we even can work at values above the presumed $\kappa_{crit}$ 
without noticeable problems.

In the computations below, we simulate the model at the values of
$\kappa$ given in Table 1. These numbers have been chosen from
preliminary results on $\kappa_{crit}(\beta)$, that have been
determined with help of restoration of PCAC along the lines of
\cite{PCACPapers}.  (The final results and their size dependence are
discussed in \cite{HiLaTe97}. It turns out that the $\kappa$-values of
Table 1 are slightly (typically 0.001) below the critical values
determined in \cite{HiLaTe97}.) These values are consistent with the
behaviour of other observables like the chiral susceptibility. 
We used these $\kappa$-values for all lattice sizes. As will
be discussed below, our results are only weakly dependent on the
precise value of the hopping parameter.
\begin{table}[htbp]
\begin{center}
\begin{tabular}{c|c|c|c|c|c|c|c|c}
$\beta$ & 0.1 & 0.5 & 1.0 & 1.5 & 2.0 & 3.0 & 4.0 & 5.0 \\
\hline 
$\kappa$ & 0.332 & 0.314 & 0.296 & 0.286 & 0.276 & 0.267 & 0.262 
& 0.260 \\
\end{tabular}
\end{center}
\caption{Values of $\kappa$ for different $\beta$.} 
\end{table}
\\
The eigenvalues of the hopping matrix $Q$ were determined by general
purpose routines for non-hermitian matrices. Since the configurations
are comparatively small, direct determination of all eigenvalues and
eigenvectors was possible throughout.

\subsection{Topological charge in the continuum and on the lattice} 

In this subsection we discuss the topological charge in the continuum and
its geometric definition on the lattice. We state the Atiyah-Singer 
Index Theorem and the less known Vanishing Theorem which holds for 
QED$_2$. 
\\

We start with collecting some results
for QED$_2$ in the continuum which can be found in the classical papers
\cite{top2}. 
There space-time is compactified on a sphere by requiring
the gauge fields to approach a pure gauge at infinity. The case of 
QED$_2$ on a (continuous) torus is discussed in \cite{Jo90}. 
In both cases one can define the topological charge
(Pontryagin index)
\begin{equation}
\nu[A] \; = \; \frac{e}{2\pi} \int d^2x \; F_{12}(x) \; , 
\label{tocha}
\end{equation}
which is an integer valued functional for classical (differentiable) 
gauge field configurations. The Atiyah Singer Index Theorem relates 
the topological charge to the index of the Dirac operator, which is given 
by the difference of the number of positive ($n_+$) and negative ($n_-$)
chirality zero modes. The
zero modes are eigenstates of the Dirac operator
\[
i \rlap{D}{\not}\;\; \; \psi(x) \; = \; 
i\gamma_\mu \left[ \partial_\mu - ie A_\mu(x) \right] \psi(x)  
\; = \; E \psi(x) \; ,
\]
with eigenvalue $E = 0$. Since $\gamma_3$ anticommutes 
with $i \rlap{D}{\not} \;\;$, the zero modes can be chosen as eigenstates 
$\psi_+ , \psi_-$ of $\gamma_3$
with $\gamma_3 \psi_+ = + \psi_+$ and $\gamma_3 \psi_-  = - \psi_-$.
$n_+$ ($n_-$) denotes the number of independent states
$\psi_+$ ($\psi_-$). The Atiyah Singer Index Theorem \cite{AtSi71} then reads
\begin{equation}
\nu[A] \; = \; n_+ \; - \;  n_- \; \; \; \; .
\label{asit}
\end{equation} 
For the case of QED$_2$ there holds another index theorem, which is 
sometimes referred to as Vanishing Theorem \cite{top2}.
\begin{eqnarray}
\nu[A] \; > \; 0 \; \; & \Longrightarrow & \; \; n_- \; = \; 0 \; ,
\nonumber \\
\nu[A] \; < \; 0 \; \; & \Longrightarrow & \; \; n_+ \; = \; 0 \; .
\label{vat}
\end{eqnarray}
The Vanishing Theorem thus states that for non-zero topological charge either 
only positive or only negative chirality eigenmodes exist.
\\

When computing the topological charge for the lattice gauge fields, we use 
the geometric definition which is based on L\"uscher's 
idea of associating a principal
bundle to each lattice configuration and defining its topological charge
through the topological charge of the bundle 
\cite{Lu82}. The case of QED$_2$ was worked out
in \cite{lattop2}. One obtains
\begin{equation}
\nu_{l}[U] \; \; = \; \; \frac{1}{2\pi} \sum_{x \in \Lambda} \theta_P(x) 
\; \; \in \; \; \mbox{Z\hspace{-1.3mm}Z} \; .
\label{toplat}
\end{equation}
The plaquette angle $\theta_P(x)$ is introduced as (compare 
(\ref{plaquel})) 
\begin{equation}
\theta_P(x) \; = \; \mbox{Im}\, \ln U_P(x) \; ,
\label{plaquang}
\end{equation}
and restricted to the principal branch $\theta_P(x) \in (-\pi,\pi)$.
Note that configurations where $U_P(x) = -1$ for some $x$ are
so-called exceptional configurations and L\"uscher's 
definition does not assign a
value $\nu_{l}[U]$ to them. (This notation should not be
confused with the widely used terminology where one calls exceptional
configurations those gauge field configurations
with bad convergence properties in the inversion algorithm 
for the Dirac operator.)
However, those configurations are of measure 
zero in the path integral. 

\section{Analytic results for the eigensystem of the fermion matrix}
This section is devoted to some analytic results on the general structure of 
the spectrum and the chiral properties of the eigenvectors
of the fermion matrix. These results are valuable tools for our
later numerical investigation of the spectrum and are necessary for a
proper interpretation of an eventual approximation of the continuum
index theorems by the lattice model.

\subsection{The general structure of the spectrum}
In this subsection we collect analytic results ({\sl S1 - S4}) 
for the spectrum of 
the fermion matrix $M$ which partly can be found in the literature 
\cite{SmVi87,ItIwYo87,WeCh79}. Here we carefully
distinguish between even and odd values of $L$ since the structure 
of the spectrum changes according to $L$ even or odd.
A proper investigation of this fact is 
necessary for understanding an eventual 
realization of the index theorems (\ref{asit}), (\ref{vat}) on the
lattice.
\\

We start with noting that the hopping matrix $Q$, defined in (\ref{Qferm}) 
is similar to its hermitian adjoint $Q^\dagger$
\begin{equation}
\Gamma_3 Q \Gamma_3 \; = \; Q^\dagger \; ,
\label{simtra}
\end{equation}
where 
\begin{equation}
\Gamma_3 = \mbox{1\hspace{-1.2mm}I}_{sites} \otimes \gamma_3
\; \; \; \; \; \; \; \; \; \mbox{with}\; \; \; \; \; \; \;
\Gamma_3^2 = \mbox{1\hspace{-1.2mm}I} \; \; \; , \; \; \; 
\Gamma_3^\dagger \; = \; \Gamma_3 \; .
\label{gamma3def}
\end{equation}
This implies that
$\alpha$ is a simultaneous eigenvalue of $Q^\dagger$ (eigenvector $v$), 
and of $Q$ (eigenvector $\Gamma_3 v$).
Let $\lambda$ be some eigenvalue of $Q$; then from 
general theorems $\overline{\lambda}$ is an eigenvalue of $Q^\dagger$
and thus also of $Q$. 
\vskip2mm 
\noindent
{\sl S1: Eigenvalues of $Q$ and hence also of $M$ (compare (\ref{Qferm}))
are either real or come in complex conjugate pairs.} 
\vskip2mm
\noindent
Furthermore for $L$ even $Q$ is also similar to $-Q$
\begin{equation}
\Xi Q \Xi \; = \; -Q \; ,
\label{qsimmq}
\end{equation}
where 
\begin{equation}
\Xi(x,y) \; = \; (-1)^{x_1 + x_2} \; \delta_{x_1,y_1} \delta_{x_2,y_2}
\otimes \mbox{1\hspace{-1.2mm}I}_{spinor} \; \; \; \; \; \mbox{with}
\; \; \; \; \; \Xi^2 \; = \; \mbox{1\hspace{-1.2mm}I} \; \; \; , \; \; \; 
\Xi^\dagger \; = \; \Xi \; . 
\label{xidef}
\end{equation}
For $L$ odd (\ref{qsimmq}) does 
not hold, since the crucial condition (compare (\ref{kron}))
\[
\Delta_{x \pm \hat{\mu}, y}
(-1)^{x_1+x_2+y_1+y_2} = - \Delta_{x \pm \hat{\mu}, y} \; ,
\]
is violated for 
$x_\mu = 1$ (or $L$). The property
$\Xi^2 = \mbox{1\hspace{-1.2mm}I}$ 
implies that if $\lambda$ is an eigenvalue of $Q$ with eigenvector
$v$, then also $-\lambda$ is an eigenvalue of $Q$ (eigenvector $\Xi v$).
Thus the spectrum obeys:
\vskip2mm
\noindent
{\sl S2: For even $L$, eigenvalues of $Q$ come in pairs $\lambda, - \lambda$.}
\vskip2mm
\noindent
Let us now turn to the eigenvalues $\mu$ of the fermion matrix $M$ itself. 
From (\ref{Qferm}) one finds: 
\vskip2mm
\noindent
{\sl S3: The eigenvalues $\mu$ of the fermion matrix 
$M$ are related to the eigenvalues $\lambda$ of the hopping matrix $Q$ through}
$\mu \; = \; 1 - \kappa \lambda \; . $
\vskip2mm
\noindent
This has the direct consequence that for any positive, real $\lambda$ there 
is a  $\kappa_0$ such that on that gauge field configuration 
$\mu(\kappa_0)=0$.

By establishing a bound on the norm of $Q$, further information on
the spectrum can be obtained. 
With (\ref{Qferm}), 
using the fact that $1 + \gamma_\nu$ and $1 - \gamma_\nu$
are proportional to orthogonal projectors
and $|U_\nu(x)| = 1$ it is straightforward to show
\[
\parallel Q_\nu \parallel_\infty \; = \; \sup_g 
\frac{\parallel Q_\nu g \parallel}{\parallel g \parallel} \; = \; 2 \; .
\]
The norm $\parallel .. \parallel$ is defined to be the $l^2$ norm 
obtained by summing over all lattice and spinor indices and $g$ is some 
(spinor) test function on the lattice. 
This implies for an eigenvalue 
$\lambda$ of $Q$ with eigenvector $v$
\[
| \lambda | \; = \; \frac{\parallel Q v \parallel}{\parallel v
\parallel} \; \leq \; \parallel Q \parallel_\infty \;  = 
\; \parallel Q_1 + Q_2 \parallel_\infty \; \leq \; 
\parallel Q_1 \parallel_\infty + \parallel Q_2 \parallel_\infty 
\; = \; 4 \; .
\]
Together with {\sl S3} one finds:
\vskip2mm
\noindent
{\sl S4: The eigenvalues $\mu$ of $M$ are 
distributed inside a circle with radius $= 4\kappa$
with center $1$ in the complex plane.} 
\vskip2mm
\noindent
The results {\sl S1 - S4} have important implications for an 
approximate realization of the index theorems (\ref{asit}), (\ref{vat}) 
on the lattice. For example a configuration with $|\nu_{l}[U]| = 1$ 
lets one expect (due to (\ref{asit}) and (\ref{vat})) exactly one 
approximate zero-mode,
i.e.~exactly one eigenvalue of $M$ with small modulus.
This eigenvalue is then necessarily real, since if
the lowest lying eigenvalue was complex, both this eigenvalue and 
its complex conjugate would have the same absolute value. The 
emergence of one small and real eigenvalue should be invariant under 
small deformations of the background field $U$.
Finally we remark, that {\sl S4} implies that for $\kappa < 1/4$
(which corresponds to bare mass $m > 0$)
exactly vanishing eigenvalues are excluded, and real eigenvalues are
positive. This corresponds to the fact that for non-vanishing 
quark masses $m$ also in the continuum exact zero modes are excluded.

\subsection{Chiral properties of the eigenvectors}
In this section we prove a theorem on the chiral properties of the 
eigenvectors of the fermion matrix. It establishes that indeed, 
as already conjectured in the end of the last paragraph, only
the eigenvectors with real eigenvalues can be interpreted as approximate
zero modes.
\\

In the continuum the zero modes can be chosen as 
chiral eigenstates, since $\gamma_3$ anticommutes with the Dirac operator. 
On the lattice the Wilson-Dirac operator has terms that explicitly
break the chiral invariance and it does not anticommute any longer
with $\gamma_3$. 
However, it is interesting to study the pseudoscalar density matrix
\begin{equation}
\chi (i,j) \; \; = \; \; ( v_i, \Gamma_3 v_j ) \; \;
\equiv \; \; v_i^\dagger \Gamma_3 v_j \; ,
\label{chidef}
\end{equation}
where $v_i, v_j$ are the right 
eigenvectors of the hopping matrix $Q$ with eigenvalues
$\lambda_i, \lambda_j$. Due to (\ref{mq}) $v_i, v_j$ are also 
eigenvectors of the fermion matrix $M$, with eigenvalues 
$\mu_i, \mu_j$ related to $\lambda_i, \lambda_j$ through {\sl S3}. 
Since $\Gamma_3$ is hermitian, we find
\begin{equation}
\chi (j,i) \; \; = \; \; \overline{\chi (i,j)} \; \; \; \; \; \; 
\mbox{and} \; \; \; \; \; \chi (i,i) \in \;
\mbox{I\hspace{-1.0mm}R} \; ,
\label{rhohermit}
\end{equation}
which shows that $\chi$ is a hermitian matrix. The following 
theorem holds:
\vskip3mm
\noindent
{\sl Theorem 1 (Vanishing entries of the chiral density matrix): } \\
\begin{equation}
\lambda_i \; \ne \; \overline{\lambda_j} \; \; \Rightarrow 
\; \; \chi(i,j) = 0 \; ,
\label{chirnondiag}
\end{equation}
and in particular for the diagonal elements
\begin{equation}
\lambda_i \; \notin \; \mbox{I\hspace{-1.0mm}R}
 \; \; \Rightarrow \; \; \chi (i,i) = 0 \; .
\label{chirdiag}
\end{equation}
Note that using {\sl S3}
the conditions $\lambda_i \; \ne \; \overline{\lambda_j}$ and 
$\lambda_i \; \notin \; \mbox{I\hspace{-1.0mm}R}$ can immediately be written 
into conditions $\mu_i \; \ne \; \overline{\mu_j}$ and 
$\mu_i \; \notin \; \mbox{I\hspace{-1.0mm}R}$ for the eigenvalues $\mu_i$
of the fermion matrix $M$. 
\vskip3mm
\noindent
{\sl Proof:}
We define the matrix $H = \Gamma_3 Q$.
From (\ref{simtra}) it follows that $H$ is hermitian and thus
$(v, H v) \in \mbox{I\hspace{-1.0mm}R}$ for arbitrary vectors $v$.
It follows (make use of the eigenvalue equation $Qv_i = \lambda_i v_i$)
\[
\mbox{I\hspace{-1.0mm}R} \; \ni \;
(v_i, H v_i) \; = \; (v_i, \Gamma_3 Q v_i) \; = \; 
\lambda_i \; (v_i, \Gamma_3 v_i) \; = \; 
\lambda_i \; \chi(i,i) \; .
\]
Since $\chi (i,i) \in \mbox{I\hspace{-1.0mm}R}$ we conclude 
that whenever $\chi (i,i)$ is non-vanishing, $\lambda_i$ has 
to be real. This proves (\ref{chirdiag}) for the diagonal 
elements. The result for the diagonal entries of the pseudoscalar density 
matrix may be found in \cite{NaNe95}. Next consider 
\begin{eqnarray}
\mbox{I\hspace{-1.0mm}R} \! \!
&  \ni &
(v_i + v_j, H [ v_i + v_j] ) =   
\lambda_i \chi (i,i) + \lambda_j \chi (j,j) +  
\lambda_i \chi (j,i) + \lambda_j \chi (i,j) \; , 
\nonumber \\ 
\mbox{I\hspace{-1.0mm}R} \! \!
&  \ni & 
(v_i + iv_j, H [ v_i + iv_j] ) =  
\lambda_i \chi (i,i) + \lambda_j \chi (j,j) -  
i \lambda_i \chi (j,i) + i \lambda_j \chi (i,j) . \nonumber 
\end{eqnarray}
Since we already showed that the first two terms on the right hand sides
are real (either $\lambda_i$ is real or $\chi (i,i)$ vanishes) the last 
equation implies (use (\ref{rhohermit}))
\begin{eqnarray}
\lambda_i \; \overline{\chi (i,j)} \; \; + 
\; \; \lambda_j \; \chi (i,j) 
& \; \in \; & \mbox{I\hspace{-1.0mm}R}\; , \nonumber \\
-\; i \lambda_i \; \overline{ \chi (i,j) } \; \; + 
\; \;  i \lambda_j \; \chi (i,j) 
& \; \in \; & \mbox{I\hspace{-1.0mm}R} \; . 
\label{realcond}
\end{eqnarray}
Introducing the abbreviations
$\chi (i,j) = a + ib , \lambda_i = x_i + iy_i$ and $\lambda_j = x_j + iy_j$ 
the vanishing of the imaginary part in (\ref{realcond}) can be 
expressed as 
\[
\left[ \matrix{ b & a \cr -a & b } \right] \; 
\left[ \matrix{x_i - x_j \cr y_i + y_j } \right]  \; \; = \; \; 
\left[ \matrix{ 0 \cr 0 } \right] \; .
\]
The matrix on the left 
has determinant $a^2 + b^2$. A non-trivial solution for the vector 
$(x_i - x_j, y_i + y_j)^T$ is possible only if this determinant vanishes. 
This implies that whenever $a^2 + b^2 \ne 0$, we must have 
\[
x_i \; = \; x_j \; \; , \; \; y_i \; =  \; - y_j \; \; 
\Leftrightarrow \; \; 
\lambda_i \; = \; \overline{\lambda_j} \; .
\]
The condition $a^2 + b^2 \ne 0$ is equivalent to $\chi (i,j) 
\ne 0$ and thus (\ref{chirnondiag}) is proven. $\Box$
\\

Theorem 1 has important consequences for an eventual interpretation 
of eigenvectors of the fermion matrix as candidates for approximate
zero modes. In particular the statement (\ref{chirdiag}) on the 
diagonal entries $\chi(i,i)$ of the pseudoscalar density matrix shows
that only eigenvectors $v_i$ with {\sl real} eigenvalues $\mu_i$
allow for non-trivial chiral properties. If one assumes that the 
index theorem of the continuum can be recovered from the continuum 
limit on the lattice, eigenvectors $v_{complex}$
with (truly) complex eigenvalues are 
already ruled out as candidates for zero modes, since they always
obey $(v_{complex}, \Gamma_3 v_{complex}) = 0$, whereas normalized
zero modes in the continuum obey $(\psi, \gamma_3 \psi)$ $=$ $\pm 1$.

\section{Numerical results for the eigensystem of the fermion matrix}
In Section 3 we explored some general aspects of the spectrum and the
chiral properties of the eigenstates of the fermion matrix using 
analytic methods. Here we numerically analyze the behaviour of the 
spectrum in Monte Carlo generated background configurations.

\subsection{Spectrum in fixed background configurations}
In this subsection we discuss the spectrum of the fermion matrix
using several {\sl fixed} background configurations. We formulate some
conjectures on the general structure of the spectrum, which will
be tested in subsequent sections using the whole Monte Carlo
ensemble of configurations.
\\

When the couplings approach their critical
values (compare Section 2.2) we find \cite{GaHiLa97c}
that the physical signature 
(spectrum of triplet and singlet currents, chiral condensates) of the 
lattice model is in  
agreement with the continuum results for QED$_2$ with {\sl massless } fermions. 
Thus we expect that the eigenvalues of the full fermion matrix (including the 
constant term) resembles the spectrum of the {\sl massless} 
continuum Dirac operator. We remark that we restrict ourselves to lattices
$L \times L$ with even $L$ in order to be able to use the maximal symmetry
including {\sl S2}. 

\begin{figure}[htbp]
\epsfysize=3.5in
\epsfbox[71 433 515 727] {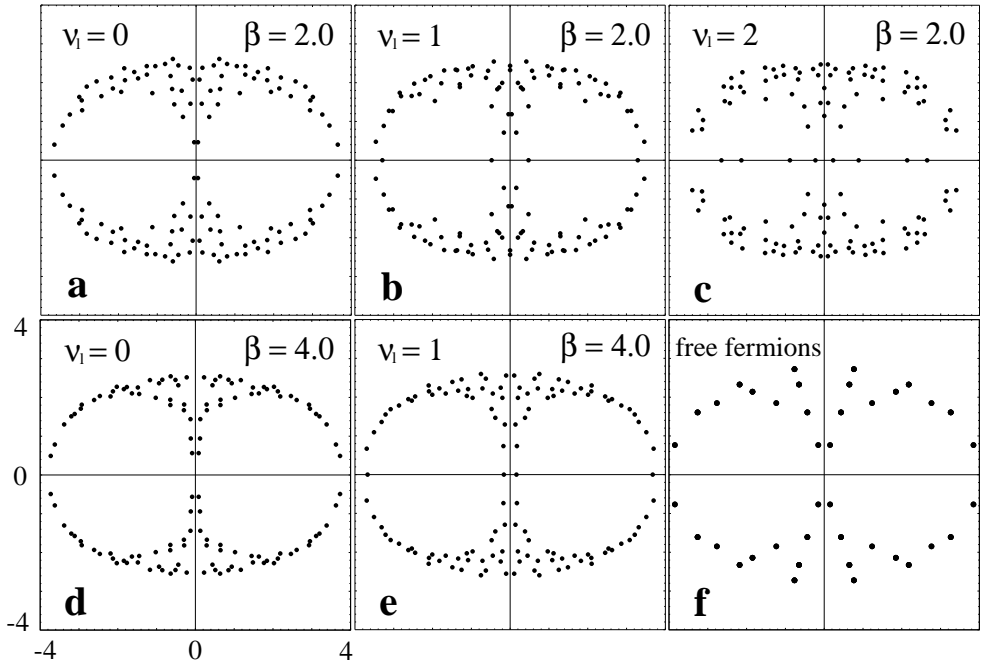}
\caption{ 
Spectrum of $Q$ in the complex plane for a $8\times8$ lattice. 
The background configurations in (a-c) were taken at 
$\beta$ = 2.0 and $\kappa$ = 0.276. Plots (a), (b) and (c)
correspond to $\nu_{l}$ = 0, 1 and 2, respectively. 
The parameters for (d) and (e) 
are $\beta$ = 4.0, $\kappa$ = 0.262, with (d) and (e) 
showing $\nu_{l} =$ 0 and 1. Finally in (f) we plot the spectrum,  
for $U=1$, $\kappa$ = 0.25 which corresponds to free massless fermions
(note that we use mixed periodic boundary conditions and that the 
eigenvalues in (f) are degenerate).
\label{specplot}}
\end{figure}

In Figures \ref{specplot}.a - \ref{specplot}.f we show the 
spectrum of the hopping matrix $Q$ for a set of `typical' 
background configurations with various values of $\nu_{l}$
(compare also \cite{Vi88,BaDuEiTh97}).
The Monte Carlo generated configurations are in equilibrium 
for the corresponding values of $\kappa$ and $\beta$.
The values of $\kappa$ were chosen near the critical $\kappa$ for 
the corresponding $\beta$.

It is obvious that all spectra obey the symmetry properties {\sl S1, S2, S4}.
The eigenvalues are either real or come in complex conjugate pairs,
they are enclosed in a circle of radius 4 around the origin in the
complex plane. Their distribution is symmetric with respect to
reflection at the imaginary axis, as expected from {\sl S2} for even $L$.

We remark, that due to {\sl S3} the largest real eigenvalues of $Q$ are to 
become the smallest real eigenvalues of $M$ and thus are the candidates for 
the approximate zero modes. 
The spectra for topological charge $\nu_{l} = 1$ 
(Figs. \ref{specplot}.b and \ref{specplot}.e) 
show one large real eigenvalue of $Q$ (corresponding to one small real
eigenvalue of $M$) as was predicted already in the discussion 
at the end of Section 3.1. This eigenvalue has the clear 
interpretation of corresponding to an approximate zero mode. 

In the sector with $\nu_{l} = 2$ (Fig. \ref{specplot}.c) the situation
is more involved. The spacing of the two largest real eigenvalues is
of similar size as their distance from the nearest pair of complex
eigenvalues. Thus the size of the corresponding eigenvalues of $M$ is
{\sl not} a proper criterion to identify them as approximate zero modes.
However from the analysis of the chiral
properties of the eigenvectors in Theorem 1 it is clear that the 2
small real eigenvalues of $M$ correspond to 
the approximate zero modes. 

It is interesting to note, that in all plots 
the number of real eigenvalues is
equal to $4\, |\nu_{l}|$. It is generally believed
\cite{ItIwYo87,Vi88,BaDuEiTh97} that the smaller real eigenvalues
are the `would be' zero modes
from the doublers, which are shifted to higher values by the Wilson
term. Thus one is tempted to conjecture (for $L$ even)
\begin{equation}
\# \; of \; real \; eigenvalues \; = \; 4 |\nu_{l}| \; .
\label{nre}
\end{equation}
However, it has to be remarked, that (\ref{nre}) cannot be exact for
arbitrary gauge field configurations. A counter example is given by
\begin{equation}
U_2(x_1,L) = - 1 \; \; \; \forall x_1 = 1, ... \; L \; \; \; \; \; 
; \; \; \; \; \; 
\mbox{all other} \; U_\mu(x_1,x_2) = 1 \; . 
\label{ucex}
\end{equation}
This configuration has $U_P(x) = 1$ for all $x \in \Lambda$ and
thus $\nu_{l}[U] = 0$. It furthermore reduces the fermion matrix
to the case of free fermions with periodic boundary conditions.
The spectrum for this case can be computed using Fourier transform
and is given by ($ n_1, n_2 = 1, ... \; L$) 
\[
1 \; - \; 2 \kappa \; \left( 
\cos\Big(\frac{2\pi n_1}{L}\Big) + 
\cos\Big(\frac{2\pi n_2}{L}\Big) \; \pm \; 
i \sqrt{2 - \cos^2\Big(\frac{2\pi n_1}{L}\Big) - 
\cos^2\Big(\frac{2\pi n_2}{L}\Big) } \right)
\; .
\]
It is obvious that the imaginary parts vanish whenever 
$n_1, n_2 \in \{L/2, L\}$.
This establishes the fact that there are real eigenvalues
although $\nu_{l}[U] = 0$, implying that (\ref{nre}) can be violated for 
certain configurations. This possible violation was also confirmed, when we 
performed a numerical analysis of the 
spectrum for randomly chosen background configurations. In this
study with static background configurations it also turned out, that
the smoother the gauge field, i.e.~plaquette variables close to 1, was chosen,
the less likely a violation of (\ref{nre}) became (although isolated configurations such as (\ref{ucex}) with all plaquette variables equal 1
still violate (\ref{nre})). We conjecture that up to isolated configurations, 
(\ref{nre}) is indeed 
correct for sufficiently smooth gauge field configurations 
(compare \cite{SmVi87,Vi88,BaDuEiTh97}). 

\subsection{The relation between the 
topological charge and the real eigenvalues}

In the last section it was 
conjectured, that (\ref{nre}) should become correct for gauge field
configurations which are smooth in some sense. In particular one
is interested in the behaviour at large $\beta$ since one wants to
construct the continuum limit when $\beta \rightarrow \infty$. It is
known that with increasing $\beta$ the gauge field becomes smoother
(plaquette variables closer to 1)
since local fluctuations are stronger suppressed by the action.
\\

\begin{figure}
\epsfysize=2.7in
\hspace*{25mm}
\epsfbox[25 165 395 465] {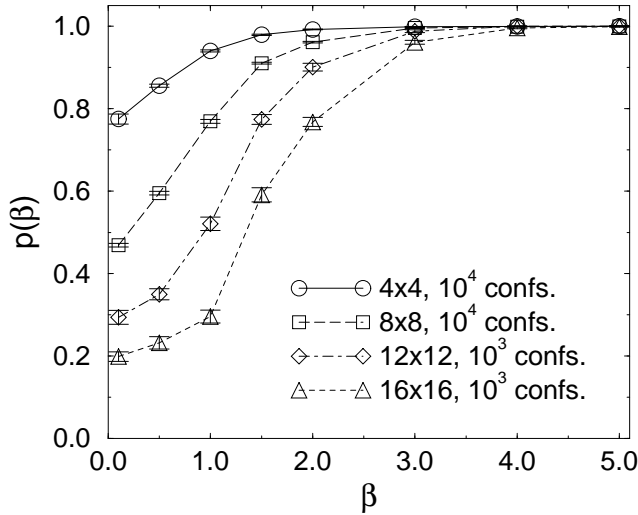}
\caption{Probability $p(\beta)$ of finding (\protect{\ref{nre}}) 
correct as a function 
of $\beta$. We show our results for lattice sizes $L=4, 8, 12$ and 16. 
The symbols are connected to guide the eye. 
\label{okvsbeta}}
\end{figure}

In order to test the conjecture that the probability $p(\beta)$
of finding (\ref{nre}) correct increases with $\beta$, 
we plot $p(\beta)$ in Fig.~\ref{okvsbeta} as a function of $\beta$ 
for various lattice sizes (L = 4,8,12, and 16). 
For each $\beta$ the value of $\kappa$
was always taken to be approximately $\kappa_{crit}$ for this $\beta$
and lattice size, as discussed in Sec.~2.2. The values of $\beta$ are
0.1, 0.5, 1.0, 1.5, 2.0, 3.0, 4.0, 5.0. For $L=4$ and $8$ we used $10^4$
configurations and $10^3$ for $L=12$ and $16$.

The figure shows clearly, that already for modestly high values of $\beta$, the
configurations obeying (\ref{nre}) entirely dominate the path integral. 
For $\beta > 3$, $p(\beta)$ is essentially equal to 1 for all lattice sizes
we analyzed. We remark, that we found only a weak dependence of 
$p(\beta)$ on $\kappa$.

What is interesting to note is the difference between the various 
lattice sizes. For fixed $\beta < 3$ the larger lattices lag behind
smaller ones in obeying (\ref{nre}). This poses an interesting question:
Do the curves for $p(\beta)$ reach a limiting curve when $L$ increases?
Ideally when performing the 
continuum limit one would first like to send $L \rightarrow \infty$ 
followed by the limit $\beta \rightarrow \infty$. If a limiting 
curve with $p(\beta) = 1$ for all $\beta$ larger than some finite 
bound $\beta_0$ emerges one could work with (\ref{nre}) for all 
$\beta > \beta_0$, independent of the lattice size. If no such limiting curve 
emerges, the bound $\beta_0(L)$ depends on $L$. However, for fixed $L$
we can use (\ref{nre}) for sufficiently large $\beta$.

\begin{figure}
\epsfysize=2.7in
\hspace*{25mm}
\epsfbox[25 165 395 465] {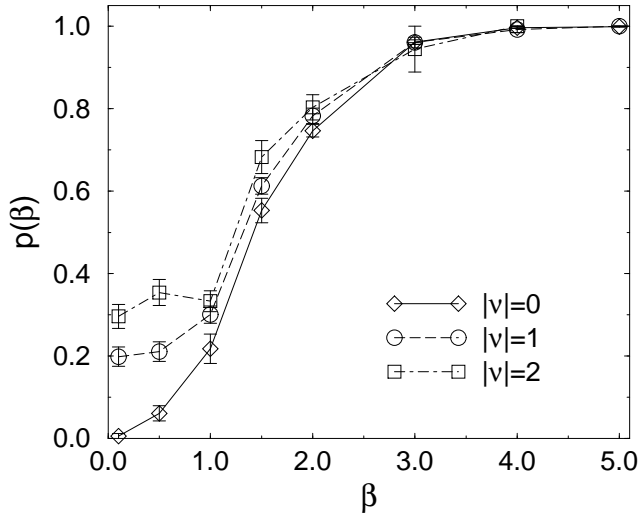}
\caption{$p(\beta)$ decomposed with respect to $|\nu_{l}|$. 
The data were computed from the configurations on a $16 \times 16$ lattice 
already used in Fig.~\protect{\ref{okvsbeta}}. The symbols are connected to 
guide the eye.
\label{okvsbetanu}}
\end{figure}

There is a second detail which deserves discussion, but is peculiar for 
QED$_2$. Although the classical index theorems (\ref{asit}), (\ref{vat}) 
are used only as a guideline 
in our decomposition of the path integral,
the distinct role of the trivial sector in the Vanishing Theorem (\ref{vat})
(which only holds for QED$_2$) should be further analyzed. The theorem 
makes no statement on the trivial sector and only
the Atiyah Singer Index Theorem (\ref{asit}) applies. Thus there is no
immediate hint, that (\ref{nre}) should also hold for $\nu_{l} = 0$. 
In order to settle this issue, we show in Fig.~\ref{okvsbetanu} our 
results for $p(\beta)$ decomposed with respect to $|\nu_{l}|$ on a 
$L = 16$ lattice. The raw data for this plot are the same as in Fig.~\ref{okvsbeta}.

We find, that for $\beta \geq 1$ the different sectors are essentially 
uniform in obeying (\ref{nre}). In particular we find no anomaly
for the trivial sector. We remark, that for larger $\beta$ and 
higher $|\nu_{l}|$ it is difficult to obtain good statistics, since
the higher sectors already have a considerably smaller weight in the
path integral (compare Section 5.1). The observed high values of
$p(\beta)$ for $\beta \leq 1$, $|\nu_{l}| \geq 1$ might be 
explained by the increase of the effective action with $|\nu_{l}|$
(see Section 5.1).
Increased values of the effective action for higher $|\nu_{l}|$ damp 
additional quantum fluctuations of the gauge field 
in the higher sectors leading to 
smoother configurations with enhanced $p(\beta)$. When $\beta$
increases, the configurations are forced to be smoother also in the 
lower sectors leading to the observed uniform behaviour in $|\nu_{l}|$
for higher $\beta$.
\goodbreak

\subsection{Properties of the spectrum in ensembles of background 
configurations}

Pursuing the program outlined in the introduction one would like to go
over from individual background configurations to an investigation of
the behaviour of the spectrum in a whole ensemble of configurations
with fixed winding number. In this section we establish some properties
of the average behaviour of the spectrum in the ensemble of the path 
integral. In particular we concentrate on the real eigenvalues which
are intimately connected to the topological charge, as established 
in the last section.
\\

\begin{figure}[htbp]
\epsfysize=1in
\epsfbox[0 488 375 567] {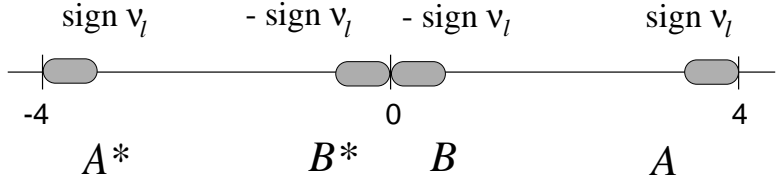}
\caption{Schematic picture of the distribution of the real eigenvalues
of the hopping matrix $Q$. The shaded areas are the domains 
$A^*, B^*, B$ and $A$ where the real eigenvalues concentrate. In the top line
we indicate how the sign of the matrix elements $\chi(i,i)$ corresponding to 
the eigenvalues in the domains is related to the sign of the 
topological charge $\nu_{l}$. \label{specschem} }
\smallskip
\end{figure}

We start with showing a schematic picture (Fig. \ref{specschem})
for the distribution of the {\sl real} eigenvalues of the hopping matrix $Q$.

The pattern displayed in Fig.~\ref{specschem} emerges clearly for smooth 
configurations and becomes more pronounced for increasing
values of $\beta$ (see also \cite{Vi88}). 
We find that the real eigenvalues are concentrated
in the small domains $A, B, A^*$ and $B^*$. The eigenvalues 
in $A$ are related to the eigenvalues in $A^*$ through the
symmetry $\lambda \leftrightarrow - \lambda$ ({\sl S2}) and the
same holds for $B$ and $B^*$. We find that the corresponding diagonal
entries of the pseudoscalar density matrix $\chi(i,i)$ are equal 
(as a matter of fact this can be proven analytically, using similar
techniques as in the proof of Theorem 1). 
Also the sign of the diagonal entries obeys a simple pattern: It coincides 
with the sign of $\nu_{l}$ for the regions $A, A^*$ and equals 
minus the sign of $\nu_{l}$ for $B$ and $B^*$. This shows that the 
eigenvectors corresponding to domain $A$, which are the approximate
zero modes (their eigenvalues of $M$ approach 0) also have the correct
chirality as expected from (\ref{asit}), (\ref{vat}). 

We find that with increasing $\beta$ the domains $A$ and $B$ shrink and
are shifted towards the limiting values 4 and 0 respectively 
(see also \cite{Vi88,BaDuEiTh97}). In 
order to quantify this statement, we computed the average values
$\overline{A}, \overline{B}$ and 
the standard deviations $\sigma_A, \sigma_B$ of the real 
eigenvalues in each of the domains $A$ and $B$. 
The eigenvalues can be attributed to $A$ or $B$ due to the chiral
properties of the eigenvectors (i.e. the sign of $\chi(i,i)$).
\begin{figure}[htbp]
\epsfysize=2.42in
\epsfbox[6 505 559 754] {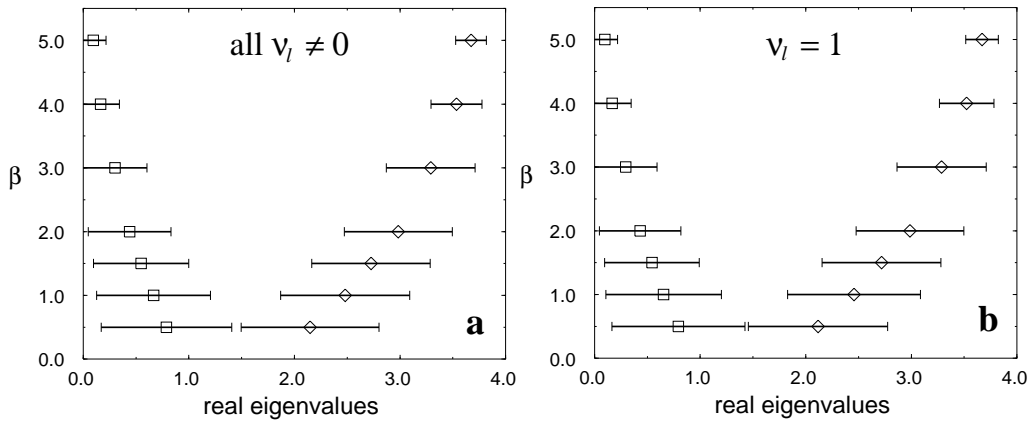}
\caption{
Average behaviour of the distribution of the real eigenvalues 
for various values of $\beta$. The data were taken on a $8 \times 8$ lattice
(see the text for statistics).
Fig.~\protect{\ref{ensemblespec}}.a (left-hand plot) 
shows the average values $\overline{A}, \overline{B}$ 
of the real eigenvalues in the domains $A$ (diamonds, right branch) and 
$B$ (squares, left branch).  
The horizontal bars show the size of the standard deviations
$\sigma_A, \sigma_B$ from the average values. 
We stress that this is not a statistical error, but gives the 
average size of the domains $A$ and $B$. In
Fig.~\protect{\ref{ensemblespec}}.b (right-hand plot) we show the 
same information but take into account only configurations with 
$\nu_{l} = 1$.
\label{ensemblespec}}
\end{figure}

Fig.~\ref{ensemblespec}.a shows our 
results for $8 \times 8$ lattices and various values of $\beta$. 
For $\beta \leq 2$ we evaluated 10$^4$ configurations in non-trivial
sectors, for $\beta = 3,4,5$ we used $10^3$ nontrivial configurations. 
The symbols give the average values $\overline{A}$ (diamonds)
and $\overline{B}$ (squares) of the real eigenvalues in 
$A, B$. horizontal bars show the corresponding standard 
deviations $\sigma_A, \sigma_B$. 
Note that this is not some kind of errorbar giving the size 
of statistical errors, but gives the average 
size of the regions $A$ and $B$. It is obvious, 
that with increasing $\beta$ the real eigenvalues get confined in 
smaller regions closer to the limits 0 and 4. 

The same picture also holds for each sector separately. 
In Fig.~\ref{ensemblespec}.b we show the same quantities as in 
Fig.~\ref{ensemblespec}.a, but restrict the ensemble to 
configurations with $\nu_{l} = 1$. Again we find that with increasing $\beta$ 
the real eigenvalues get confined in smaller regions closer to the limits 
0 and 4. 
\begin{figure}[htbp]
\epsfysize=2.3in
\hspace*{31mm}
\epsfbox[25 165 405 475] {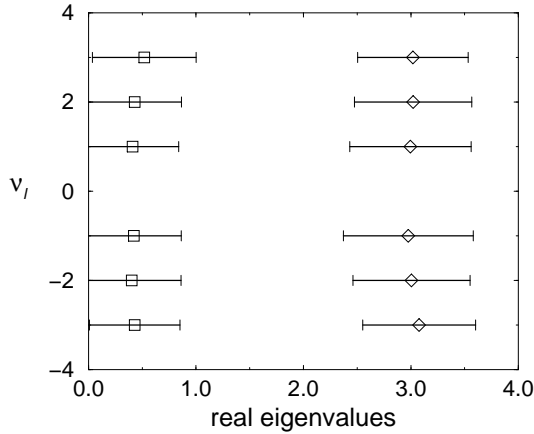}
\caption{
Average behaviour of the distribution of the real eigenvalues 
for different $\nu_{l}$. The data were computed from 2300 configurations
in non-trivial sectors on a $16 \times 16$ lattice 
at $\beta = 2, \kappa = 0.276$. 
The symbols give the average values $\overline{A}, 
\overline{B}$ of the real 
eigenvalues in $A$ (diamonds, right-hand branch) and 
$B$ (squares, left-hand branch). The horizontal bars are the
corresponding standard deviations $\sigma_A, \sigma_B$. 
\label{nucompare}}
\end{figure}

If one compares the different topological sectors ($\beta$ fixed) 
one finds that the average 
values $\overline{A}, \overline{B}$ 
of the real eigenvalues in $A, B$ are rather independent of 
$\nu_{l}$ and also the size of the domains $A, B$ is rather 
invariant. This is demonstrated in 
Fig.~\ref{nucompare} where we show average value and standard deviation 
of the eigenvalues in $A$ and $B$ as a function of the topological charge.
We used 2300 configurations in non-trivial sectors, generated on a 
$16 \times 16$ lattice at $\beta = 2, \kappa = 0.276 \;
(\approx \kappa_{crit})$.
\begin{figure}[htbp]
\epsfysize=2.3in
\hspace*{31mm}
\epsfbox[25 165 405 475] {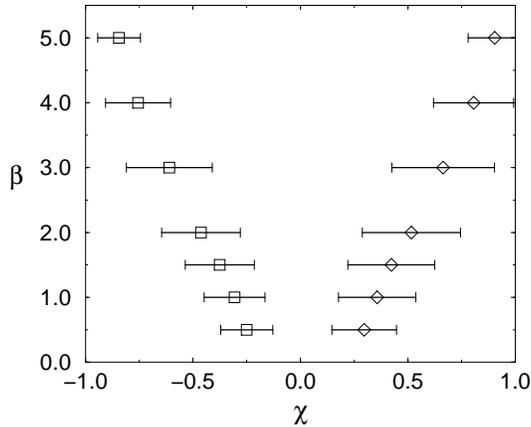}
\caption{
Average values $\overline{\chi_A}, \overline{\chi_B}$ of the entries 
$\chi(i,i)$ of the pseudoscalar density matrix corresponding to eigenvalues
in $A$ (diamonds, right branch) and $B$ (squares, left branch) 
in the $\nu_{l} = 1$ sector. The horizontal bars show the size 
of the standard deviation $\sigma_\chi$ from the average value.
The data were computed from the same configurations used in 
Fig.~\protect{\ref{ensemblespec}}.
\label{ensemblechi}}
\end{figure}

Also the diagonal entries $\chi(i,i)$ of the pseudoscalar density matrix show
a simple behaviour, monotone in $\beta$. We computed the average values
$\overline{\chi_A}, \overline{\chi_B}$ of the $\chi(i,i)$ 
and their standard deviation 
$\sigma_\chi$ from this value for each
domain $A, B$ separately. Fig.~\ref{ensemblechi} shows our result for the same
sample used in Fig.~\ref{ensemblespec}.b. $\nu_{l}$ is positive (=1), and
as expected from (\ref{asit}) and (\ref{vat}) domain $A$ 
(diamonds) has positive $\chi(i,i)$, while in region $B$ (squares)
we find negative values. Both branches show a clear trend towards
the continuum values $-1$ and 1 with increasing $\beta$. Also the interval
containing $\chi(i,i)$ (horizontal bars) shrinks for larger $\beta$. 
\\

We summarize the emerging picture for the structure of the spectrum
in the ensemble of the path integral as follows ($L$ is even):
\vskip4mm
\noindent
{\sl 
E1: Only real eigenvalues can correspond to approximate zero modes,
since for complex eigenvalues the diagonal entries of the pseudoscalar
density matrix vanish identically (Theorem 1).
\vskip2mm
\noindent
E2: For large $\beta$ the path integral is entirely dominated
by configurations obeying {\sl \# of real eigenvalues = $4 |\nu_{l}|$}
(Fig. \ref{okvsbeta}).
\vskip2mm
\noindent
E3: The real eigenvalues group according to the scheme depicted
in Fig. \ref{specschem}. The domains $A, B, A^*, B^*$ are related
through {\sl S2} and the doubler symmetry.
\vskip2mm
\noindent
E4: With increasing $\beta$ the domains $A$ and $B$ shrink and move towards
the limiting values 4 and 0 (Fig.~\ref{ensemblespec}). 
\vskip2mm
\noindent
E5: 
The sign of the diagonal entries $\chi(i,i)$ of the pseudoscalar
density matrix coincides with the sign of $\nu_{l}$ for the eigenvalues in 
$A, A^*$ (as expected from (\ref{asit}), (\ref{vat}))
and is minus this sign for $B, B^*$. With increasing $\beta$,
the values of $\chi(i,i)$ get confined in shrinking regions approaching 
the limiting  values $\pm 1$
(Fig. \ref{ensemblechi}).
\vskip2mm
\noindent
E6: The behaviour of the eigensystem is essentially uniform in $\nu_{l}$
(Figures \ref{okvsbetanu} and \ref{nucompare}).
} 
\vskip4mm

In the remaining sections we use these results for analyzing 
the contribution of different topological
sectors to the path integral and to vacuum expectation values
of various operators.

\section{Applications}

In this section we bring in the harvest from the 
previous investigation of the interplay between topological charge and
behaviour of the spectrum of the fermion matrix. 
We will demonstrate, that the analytic results {\sl S1 - S4} and 
Theorem 1, together with the results {\sl E1 - E6} on the
spectrum in ensembles of background configurations provide 
a powerful tool for analyzing physical questions in lattice gauge theory.
Here we concentrate on local and bulk quantities. The study of the 
dependence of the spectrum (which requires the computationally more 
demanding analysis of two-point functions (see also \cite{Di95})) 
on the topological charge is reserved for a
forthcoming publication \cite{GaHiLa97c}.

\subsection{Effective action and fermion determinant}
In this first application we concentrate on the behaviour of the
effective action and in particular of the fermion determinant in 
the various sectors.
\\

After integrating out the fermions (note that we use 2 explicit 
flavors of fermions (not to be mixed up with the doublers 
implicit in the Wilson action) one obtains the effective action for 
the gauge fields
\begin{equation}
S^{eff}[U] \; \; = \; \; S_g[U] \; + \; S_f^{eff}[U] \; \;
= \; \; S_g[U] \; - \; \ln \det M[U]^2 \; . 
\label{seff}
\end{equation}
It is interesting to analyze typical values of the effective action in 
various topological sectors. For the gauge field part of the action it
is known \cite{BePoSchTy75}, that there exist lower bounds of the 
action for each sector. In a previous article \cite{GaHiLa97a} we 
analyzed the role of these bounds in the case of pure U(1)$_2$ gauge
theory on the lattice (we will comment on this below). Due to the relation 
of topological charge and zero modes via the index theorems, also the fermion 
determinant depends on the topological sectors in a non-trivial way.

\begin{figure}[htbp]
\epsfysize=2.1in
\epsfbox[3 521 601 754] {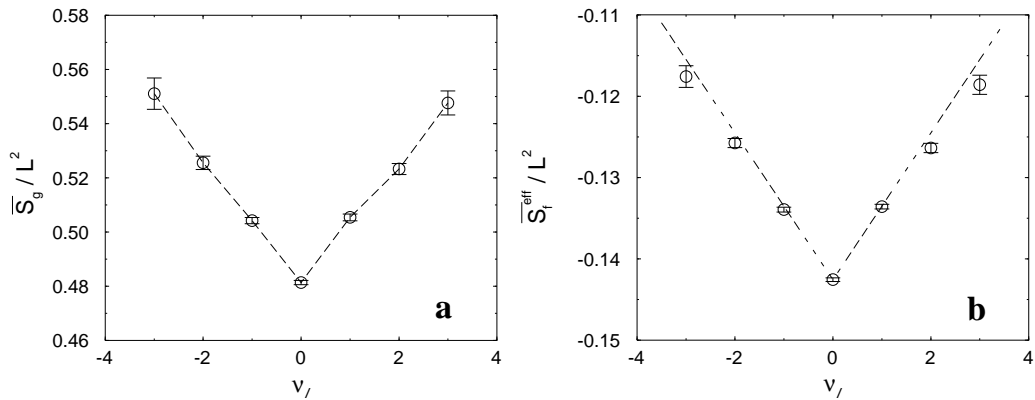}
\caption{Average value of the contributions of the gauge field action 
$\overline{S}_g(\nu_{l})/L^2$ ((a) left-hand side) and of the fermions 
$\overline{S}_f^{eff}(\nu_{l})/L^2$ ((b) right-hand side) to the 
effective action. The average was taken for each topological sector 
separately. The data were obtained from a simulation on a $16 \times 16$
lattice at $\beta = 2, \kappa = 0.276$ using $10^4$ configurations.
In (a) we connect the symbols to guide the eye.
The straight line in (b) was drawn 
using (\protect{\ref{seffp}}) below.\label{seffvsnu}}
\end{figure}

In Fig.~\ref{seffvsnu} we show our results for the averages of the gauge field 
action $\overline{S}_g(\nu)$ and the effective action from the 
fermions $\overline{S}^{eff}_f(\nu)$ in each of the sectors. 
We give the 
results per plaquette, i.e.~normalized with $L^{-2}$. 

Fig.~\ref{seffvsnu}.a (left-hand plot) shows the average value $\overline{S}_g(\nu)$
of the gauge field action in each topological sector separately.
We find a behaviour which is essentially 
linear in $|\nu_{l}|$. If one compares this
result with the pure gauge theory \cite{GaHiLa97a} the linear behaviour is 
rather surprising. In the quenched case we found a behaviour of the gauge field
action which is quadratic in $\nu_{l}$ with a variation in  
$\nu_{l}$ which is of one order of magnitude smaller than here.
This result was 
understood in a quantitative way using an argument based on the topology
of the configurations \cite{GaHiLa97a}. The essentially linear behaviour 
(plus eventually a small quadratic term) in 
Fig.~\ref{seffvsnu}.a has no such simple explanation. In the unquenched 
model it is not the gauge field action alone which is minimized, but the 
effective action (\ref{seff}) which is the sum of the gauge field action and 
the contribution from the fermion determinant. Obviously the influence of
the fermions adds a linear term which entirely changes
the picture of the pure gauge theory. The 
interplay of the two terms in (\ref{seff}) does not allow for a simple topological explanation as in \cite{GaHiLa97a}. 

There is however a lesson from the study of the pure gauge theory 
\cite{GaHiLa97a} which is also confirmed in the unquenched model:
The lower bounds for the action in each topological sector are at 
least one order of magnitude smaller than the average value of the 
gauge field action. We also computed the distribution of the values 
of the action, and find that quantum 
fluctuations keep most of the configurations high above the lower 
bounds in each sector \cite{GaHiLa97a}. The same observation is true
for the model with fermions. This is in particular remarkable,
since the lower bounds given in \cite{GaHiLa97a} are saturated by 
classical configurations with constant electric field, carrying topological 
charge. The path integral for the Schwinger model on the 
torus including all topological sectors can be constructed by dressing
these classical configurations with quantum fluctuations \cite{Jo90}.
The observed high values of the gauge field action for the main part of the 
configurations is a demonstration of the importance of quantum
fluctuations (see \cite{GaHiLa97a} for a more detailed discussion).
\\

Fig.~\ref{seffvsnu}.b (right-hand plot) shows the average value 
$\overline{S}_f^{eff}(\nu_{l})/L^2$ of the 
contribution of the fermions to the effective action for each
topological sector. Again we find a behaviour essentially linear in $|\nu_{l}|$.
This behaviour can be understood in a quantitative way using the results 
from the previous sections, as will be shown now. 

From {\sl S2, S3} it follows ($L$ even) that real eigenvalues 
$\rho$ of $M$ come in pairs 
\begin{equation}
\rho^{(1)} \; = \; 1 - \kappa \; w \; \; \; , \; \; \;
\rho^{(2)} \; = \; 1 + \kappa \; w \; ,
\label{realpair}
\end{equation}
with $0 \leq w \leq 4$. {\sl E2} and {\sl E3} establish, that for large 
enough $\beta$
\begin{equation}
\# \; of \; real \; pairs \; \; = \; \; 2 |\nu_{l}| \; , 
\label{nrp}
\end{equation}
and that the real pairs are equally distributed among the domains 
$A \cup A^*$ and $B \cup B^*$. The determinant of $M$ is the product of 
all eigenvalues. For the contribution of the real pair (\ref{realpair})
one obtains 
\begin{equation}
R \; \; = \; \; \rho^{(1)}\, \rho^{(2)} \; \; = \; \; 
1 \; - \; (\kappa w)^2 \; .
\label{realfact}
\end{equation}
In {\sl E4} we establish, that the regions $A$ and $B$ (and of course also
their {\sl S2} images $A^*$ and $B^*$) shrink and approach their limiting
values 0 and $\pm 4$). Thus we obtain characteristic contributions from the 
real eigenvalues in $A \cup A^*$ and $B \cup B^*$ 
described by the average values $\overline{A}$, $\overline{B}$ 
\begin{equation}
R_A \; = \; 1 - (\kappa \overline{A})^2 \; \; \; \; \;  , \; \; \; \; \; 
R_B \; = \; 1 - (\kappa \overline{B})^2 \; .
\end{equation}
The contribution of the complex eigenvalues is more difficult to handle.
However by inspecting the spectra for several configurations
(see Fig.~\ref{specplot} for a few examples) we find that whenever 
$|\nu_{l}|$ increases by one unit, one of the complex quadruplets vanishes 
and is turned into two real pairs. The rest of the complex eigenvalues remains
rather unperturbed giving some positive (note that complex
eigenvalues come in complex conjugate pairs) constant $c$. 
As long as $|\nu_{l}|$ is considerably smaller than the 
overall amount of eigenvalues $2L^2$, we make the following ansatz 
for the contribution of the fermion determinant to the effective action:
\begin{equation}
S^{eff}_f \; = \; - \ln \det(M)^2 \; \simeq \;
- \ln \Big( c  ( R_A R_B )^{|\nu_{l}|} \Big)^2 \; \; = \; \; 
C  \; - \; |\nu_{l}| \; 2 \ln |R_A R_B| \; .
\label{seffp}
\end{equation}
The expression gives a behaviour linear in $|\nu_{l}|$ and predicts also the
slope of the line. The constant term has to be taken from the numerical 
data. Evaluating 2300 configurations for these parameters ($L=16, \beta =2,
\kappa = 0.276$) we obtained $\overline{A} = 2.996(8)$,
$\overline{B} = 0.421(7)$.
The straight line in Fig.~\ref{seffvsnu}.b
was drawn using (\ref{seffp}) with these numbers. We find good agreement
with the Monte Carlo data. Eq. (\ref{seffp}) slightly overestimates 
the Monte Carlo results for larger $|\nu_{l}|$. This might be due to the fact
that by increasing $|\nu_{l}|$ by one, four new real eigenvalues emerge causing 
one complex quadruplet to vanish. Thus the contribution from the complex 
eigenvalues is not entirely independent of $|\nu|$ as we assumed in our 
ansatz. This effect should become smaller with increasing $L$. 
We believe that the essentially linear relation, and the good performance
of (\ref{seffp})  shows that the behaviour of the fermion 
determinant as a function of $\nu_{l}$ can be understood from 
the analysis of the spectrum of the fermion matrix in Sections 3 and 4 in 
a quantitative way (compare also \cite{BaDuEiTh97}).  
\\

We end this section with a remark on the positivity of the fermion 
determinant $\det M$. From {\sl S1 - S3} we know that the 
complex eigenvalues come in quadruplets
\begin{eqnarray}
\mu^{(1)} \; = \; 1 - \kappa \; (x + iy) \; \; \; & , & \; \; \; 
\mu^{(2)} \; = \; 1 - \kappa \; (x - iy) \; , \nonumber \\ 
\mu^{(3)} \; = \; 1 + \kappa \; (x + iy) \; \; \; & , & \; \; \; 
\mu^{(4)} \; = \; 1 + \kappa \; (x - iy) \; , \nonumber 
\end{eqnarray}
contributing a factor
\[
\; \prod_{\alpha =  1}^4 \mu^{(\alpha)} \; = \; 
\Big[ (1 - \kappa x)^2 + \kappa^2 y^2 \Big]
\Big[ (1 + \kappa x)^2 + \kappa^2 y^2 \Big] \; ,
\]
to the fermion determinant. Obviously 
the factor from a complex quadruplet is strictly positive.  

The second order polynomial $1 - (\kappa\,w)^2$ 
which one obtains for the 
contribution of a pair of real eigenvalues (\ref{realfact})
becomes negative for  $\kappa\,w > 1$.
Thus a negative value of
the fermion determinant can only come from pairs of real eigenvalues.
Since the emergence of real eigenvalues is connected
to gauge field configurations with non vanishing topological charge
(compare  {\sl E2}) the positivity properties of the determinant
are as well. In the literature one often encounters the approximation
\[
\det(M) \; \sim \;  |\det(M)| \; ,
\]
used to simulate odd numbers of flavors. This might possibly lead to 
a wrong weight from non-trivial topological sectors.

In order to investigate this possibility we performed several simulations
on $L = 8$ lattices at $\beta = 2$. The values of $\kappa$ where chosen above 
the critical $\kappa$ (which is $\kappa_{crit} \approx 0.276$ for $\beta$ = 2)
in order to force the determinant to negative 
values. $\kappa$ varied between 0.280 and 0.335. For $\kappa \geq 0.320$
we found negative values of the determinant in the non-trivial sectors. 
It is interesting to remark, that we then also found a considerably enhanced
abundance of configurations in the higher sectors. The mechanism for this
trapping is due to the contribution of the real eigenvalues to the effective
fermion action, which is essentially the sum of the logarithms 
of the eigenvalues squared (note that we squared the determinant 
for the updating). Once one of the real eigenvalues assumes a negative value
(thus causing the determinant to be negative), it can {\sl continuously} go 
back to a positive value only by passing through zero. This however 
drives the logarithm and thus the effective action to infinity.
The only way back 
is through a discontinuous jump, which probably costs a big amount of 
action. This creates a barrier, which traps the system in a topologically
non-trivial sector.
\begin{figure}[htbp]
\epsfysize=2.7in
\hspace*{23mm}
\epsfbox[30 181 407 466] {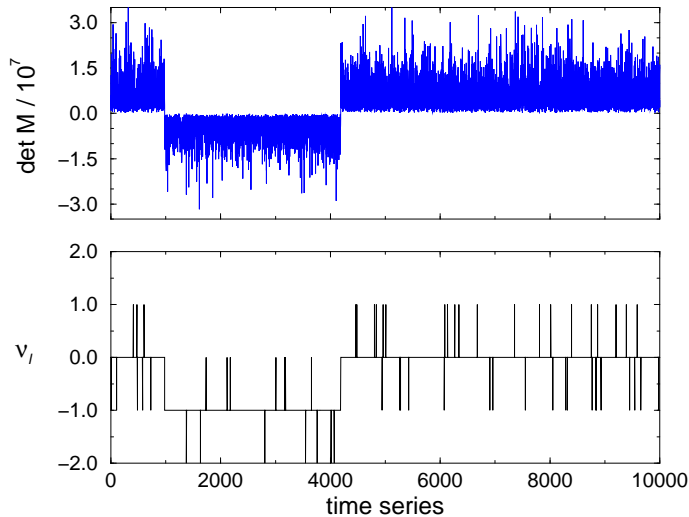}
\caption{
Time series of the fermion determinant and the topological charge.
\label{times}}
\end{figure}
 
Fig.~\ref{times} shows a time series for the determinant and
the topological charge from $10^4$ configurations at $\beta = 2$,
$\kappa = 0.33$. For approximately
the first 1000 configurations the topological charge is mainly zero
with a few configurations that fluctuate to $\nu_{l} = \pm 1$. The determinant
has only positive values. Then for approximately 3400 configurations,
the determinant has negative values, and the topological charge
is mainly trapped at $\nu_{l} = -1$ implying that the fermion matrix
has 4 real eigenvalues (and thus can have negative values). 
There are a few configurations with $\nu_{l} = 0$ or 2.
From the sign of the determinant we conclude that there still are
quadruplets of real eigenvalues, although e.g. $\nu_{l}$ vanishes.
These are just the configurations violating (\ref{nre})
in agreement with our observation in Fig.~\ref{okvsbeta}.
However, the system is obviously trapped in the 
$\nu_{l} = -1$ sector for more than 3000 configurations. After that a
tunneling back to positive values of the determinant occurs and the 
system evolves again with positive values of the determinant and
the distribution of the values of $\nu_{l}$ is again symmetric around 0.

\subsection{The pseudoscalar density}

In this section we discuss the contributions of different topological sectors
to the vacuum expectation value of the pseudoscalar density. This operator 
plays an important role in the so called `fermionic definition' 
\cite{KaSeSt86} - \cite{ItIwYo87} of the 
topological charge which uses the chiral Ward identity to define the
topological charge and is also used in the lattice derivation of the 
Witten-Veneziano formula \cite{SmVi87b}.
\\

\begin{figure}[htbp]
\epsfysize=2.7in
\hspace*{24mm}
\epsfbox[70 418 472 729] {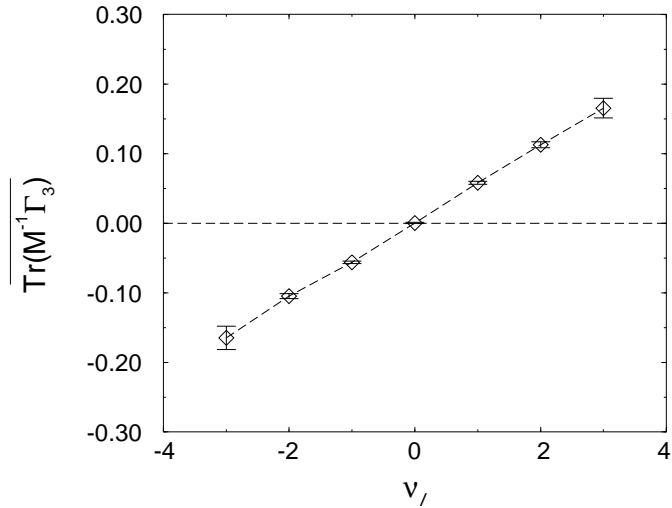}
\caption{
Average values of $\overline{\mbox{Tr}( M^{-1} \Gamma_3 )}$
in different topological sectors. The symbols show the results
from the Monte Carlo simulation ($L = 16, \beta = 2, \kappa = 0.276, 10^4$ configurations). The symbols are connected to guide the eye. 
\label{pdplot}}
\end{figure}

Due to translation invariance, the vacuum expectation value of the
pseudoscalar density can be written as
\begin{equation}
\langle \overline{\psi}(x) \gamma_3 \psi(x) \rangle \; = \; 
\frac{1}{L^2} \sum_{x \in \Lambda} 
\langle \overline{\psi}(x) \gamma_3 \psi(x) \rangle \; = \;
\frac{1}{Z \; L^2} \int [dU] e^{-S^{eff}[U]} 
\; \mbox{Tr} \; \Big( M^{-1} \Gamma_3 ) \; .
\end{equation}
In Fig.~\ref{pdplot} we show our results for the average 
value $\overline{\mbox{Tr}( M^{-1} \Gamma_3 )}$ in ensembles of background 
configurations with different $\nu_{l}$
($L = 16, \beta = 2, \kappa = 0.276, 10^4$ configurations). 
Note that in these averages, the 
observable was not multiplied with the Boltzmann factor 
$\exp( - S^{eff}[U] )$. 

The Monte Carlo data show a linear behaviour in $\nu_{l}$ with a positive 
slope. Such a behaviour can already be expected from the spectral decomposition
of the continuum Dirac operator in classical background configurations. 
The continuum operator 
$i \rlap{D}{\not}\;\;$ is hermitian, implying that eigenstates to 
different eigenvalues are orthogonal. If $\psi$ is an eigenstate with 
eigenvalue $E$, then due to $\{i \rlap{D}{\not}\;\;, \gamma_3 \} = 0$,
$\gamma_3 \psi$ has eigenvalue $-E$. Thus the matrix elements 
$< \psi | \gamma_3 | \psi > $ vanish whenever $\psi$ is not a zero 
mode. The only states that can contribute in the spectral decomposition of 
the pseudoscalar density are the zero modes. Since their
number is equal to $\nu$ due to Index and Vanishing Theorem (\ref{asit}),
(\ref{vat}) the linear behaviour follows for classical background 
configurations. 

The lattice Wilson-Dirac operator is not hermitian, even worse, not
normal. 
Thus its (right) eigenvectors do not form an $orthonormal$
basis, and the continuum argument cannot be taken over to the lattice.
However, we established ({\sl E1}) that the real eigenvalues on the lattice
take over the role of the zero eigenvalues in the continuum. We 
furthermore demonstrated ({\sl E2}) that the number of real eigenvalues is
proportional to $|\nu_{l}|$. Finally we found ({\sl E5})
that the chiral properties
(i.e.~the sign of the entry in the pseudoscalar density matrix)
of the real eigenvalues in the physical region $A$
behaves as in the continuum. These three ingredients allow to understand the 
observed linear behaviour of the pseudoscalar density.

\section{Discussion}
In this study we have demonstrated, that the existence of an integer
valued lattice equivalent of the topological charge has important 
consequences. We believe, that the use of the topological charge on 
the lattice goes far beyond its applications in continuum formulations
of gauge theories. We have shown, that a careful analytic investigation
of the spectral decomposition of the fermion matrix supports the
interpretation of the real eigenvalues (and the corresponding 
eigenvectors) as the zero eigenvalues (zero modes) of 
the continuum Dirac operator. It should be stressed again, that this is a 
result for {\sl all} configurations contributing in the continuum limit, while
the Index Theorems in the continuum only hold for classical
gauge fields. When concentrating only on the real 
eigenvalues (which is much less demanding than treating the whole spectrum), numerical methods were successfully used to analyze their distribution
when approaching the continuum limit. We have shown, that the results
for the spectrum can be used to understand the interplay between
topological charge and physical quantities in a quantitative way. 

As discussed in the introduction, QED$_2$ has many features in common with
QCD$_4$. The physical picture we obtained for the 2-dimensional theory
thus provides an interesting model for the 4-dimensional case. We remark,
that not only the physical picture, but also some of the mathematical 
results we obtained may be extended to QCD$_4$. In particular  
Theorem 1 which is the conceptual backbone of this study has
a straightforward generalization to the 4-dimensional case \cite{Ga97}.

Other aspects are certainly considerably simpler in the 2-dimensional 
world. In 4-dimensions and for non-abelian gauge groups, the fermion
matrix has much larger dimension, and explicit diagonalization becomes
a challenging enterprise \cite{SeDaBa88}. However, with the identification
of the real part of the spectrum as the trace of non-trivial topological
charge, related physical questions might be tackled without knowing 
all of the spectrum. When concentrating only on the real part, 
computational simplifications might be possible \cite{Ga97}.

Although from a conceptual point of view most satisfactory,
the geometric definition of the topological charge is very costly
to evaluate for non-abelian gauge groups. Other definitions such as the 
`fermionic approach' using the chiral Ward identity 
\cite{KaSeSt86} - \cite{ItIwYo87} might be simpler to compute. 
The latter approach essentially
makes use of the pseudoscalar density operator. We expect that also the 
results for the pseudoscalar density in Section 5.2 can be 
taken over to QCD$_4$ \cite{Ga97} and might lead to a considerable
simplification of the fermionic ansatz for the topological charge. 
\\
\\
{\bf Acknowledgment:}
We thank Philippe de Forcrand, Helmut Gausterer,
Erhard Seiler, Gordon Semenoff and Renate Teppner 
for remarks and interesting discussions.

\end{document}